\documentclass[prd,a4paper,nofootinbib,
showpacs, 
]{revtex4}
\usepackage{graphicx}
\def\eq#1{{eq.~(\ref{#1})}}

\def\Tr{\mbox{Tr}\,}

\def\hbar{\hspace{0pt}\raisebox{1pt}{$-$} \hspace{-7pt} h}

\def\5{\overline 5}

\newcommand{\be}{\begin{equation}}
\newcommand{\ee}{\end{equation}}
\newcommand{\bea}{\begin{eqnarray}}
\newcommand{\eea}{\end{eqnarray}}
\newcommand{\nn}{\nonumber}

\begin{document}
\title[LHC]{Possible experimental signatures at the LHC \\
of strongly interacting electro-weak symmetry breaking }
\date{\today
}
\author{M.~Fabbrichesi}
\author{L.~Vecchi}
\affiliation{INFN, Sezione di Trieste and\\
Scuola Internazionale Superiore di Studi Avanzati\\
via Beirut 4, I-34014 Trieste, Italy}
\begin{abstract}
\noindent 
If electro-weak symmetry is broken by a new strongly interacting sector, new physics will probably manifest itself in gauge boson scattering at the LHC. The relevant dynamics is well described in terms of an effective lagrangian. We discuss the probable size of the coefficients of the relevant operators under a combination of model-independent constraints and reasonable assumptions based on two models of the strongly interacting sector. We compare these values with LHC sensitivity and argue that they will be too small to be seen.  Therefore, the presence of vector and scalar resonances required by unitarity will be the only characteristic signature. We analyze the most likely masses and widths of these  resonances.
\end{abstract}
\maketitle
%
\section{Motivations} 
\label{sec:mot}
If the breaking of the electro-weak symmetry is due to a new and strongly interacting sector, it is quite possible that the LHC will not discover any new fundamental particle below the scale of 2 TeV. In this scenario---in which there is no SUSY and no light (fundamental or composite) Higgs boson to be seen---it becomes particularly relevant to analyze the physics of gauge boson scattering---$WW$, $WZ$ and $ZZ$---because it is here that  the strongly interacting sector should manifest itself most directly. 

Gauge boson scattering  in this regime looks  similar in many ways to $\pi\pi$ scattering in QCD and similar techniques can be used.
The natural language  is that of the effective electro-weak lagrangian introduced in \cite{eff-lag}. This lagrangian contains all dimension four operators for the propagation and interaction of the Goldstone bosons of the breaking of the global $SU(2)\times U(1)$ symmetry. If we knew the coefficients of these operators we could predict the physics of gauge boson scattering at the LHC. Unfortunately the  crucial coefficients  do not enter directly in currently measured observables. We do not know their values   and constraints on them can only be inferred by their effect in small loop corrections to the EW observables. Accordingly they are rather weak. In addition,  even though the LHC will explore these terms directly, its sensitivity is not as good as we would like it to be and an important range of  values will remain unexplored.

This lack of predictive power can  be ameliorated if we assume  some model of the strong dynamics responsible of the electro-weak symmetry breaking. In this case, additional relations among the coefficients can be found and used to relate them to known constraints.
Our strategy is therefore to use our prejudices---that is,  model-dependent relationships among the coefficients of the effective lagrangian---plus general constraints coming from causality and analyticity of the amplitudes to see what values the relevant coefficients of the effective electro-weak lagrangian  can assume without violating any of the current bounds. 

We are aware that in many models the relations among the coefficients we utilize  can be made weaker and therefore our bounds will not apply. Nevertheless we find it useful to be as conservative as possible and explore---given what we know from electro-weak precision measurements and taking the models at their face values---what can be said about gauge boson scattering if  electro-weak symmetry is  broken by a strongly interacting sector.
Within this framework, we find that the crucial coefficients  are bound to be  smaller than the expected sensitivity of the LHC and therefore  they will be probably not be detected directly.

This is  not the end of the story though. The cutoff scale of the effective theory is given by the energy at which unitarity is lost. This is around 1.3 TeV in the case of the electro-weak theory as described by the effective lagrangian at the tree level. Unitarity is recovered after introducing additional states which are the Higgs boson in the case of the standard model while they are resonances made of bound states of the strongly interacting sector in our case.
On a more practical level,  there exist unitarization procedures that move the scale at which unitarity is lost  to higher values and we will consider one of them. It is characteristic  of these procedures to automatically include the necessary resonances in the spectrum.  The presence of  resonances is particularly interesting if the coefficients of the effective lagrangian cannot be measured. They may well be the only signatures of the strongly interacting sector accessible at the LHC. We discuss in same detail the most likely masses  and widths of these resonances and their experimental signatures.

\vskip1.5em
\section{Gauge boson scattering} 
\label{sec:con}
Consider the case in which the LHC will not find any  new particle propagating under an energy scale $\Lambda$ around 2 TeV. By new we mean those particles, including  the scalar Higgs boson, not directly observed yet.
In this case---since $\Lambda\gg m_W$---the physics of gauge boson scattering is well described by the standard model (SM) with the addition of the effective  lagrangian containing all the possible electro-weak (EW) operators for the
Goldstone bosons (GB)---$\pi^a$, with $a=1,2,3$---associated to 
the $SU(2)_L\times U(1)_Y\rightarrow U(1)_{em}$ symmetry breaking.
The GB are written as an $SU(2)$ matrix
\be
U=\exp\left(i\pi^a\sigma^a/v\right)\, ,
\ee
where $\sigma^a$ are the Pauli matrices and  $v=246\;\hbox{GeV}$ is the electro-weak vacuum.
The GB couple to the EW gauge and fermion
fields in an $SU(2)_L \times U(1)_Y$ invariant way.
As usual, under a local $SU(2)_L\times U(1)_Y$ transformation $U\rightarrow L U R^\dagger$, with $L$ and $R$ an $SU(2)_L$ and $U(1)_Y$ transformation respectively. The EW precision tests require an approximate $SU(2)_C$ custodial symmetry to be preserved and therefore we assume $R \subset SU(2)_R$.

The most general lagrangian respecting the above symmetries, together with $C$ and $P$ invariance, and up to dimension 4 operators is  given in the references in~\cite{eff-lag} of which we mostly follow the notation: 
\begin{eqnarray} 
 {\cal L} &=& 
 \frac{v^2}{4} \Tr [(D_{\mu}U)^\dagger(D^{\mu}U)] + \frac{1}{4}a_0g^2v^2[Tr(TV_\mu)]^2 
                + \frac{1}{2}a_{1}g g' B_{\mu\nu}Tr(T W^{\mu\nu}) \nn \\
                & + &\frac{1}{2}i a_{2}g' B_{\mu\nu} Tr(T \left[ V^{\mu},V^{\nu}\right]) + ia_{3}gTr(W_{\mu\nu}[V^{\mu},V^{\nu}]) \nn \\ 
               &+& a_{4} [ Tr(V_{\mu}V_{\nu})] ^2 + a_{5} [ Tr(V_{\mu}V^{\mu})] ^2 + a_6Tr(V_\mu V_\nu)Tr(TV^\mu)Tr(TV^\nu) \nn\\ 
               &+& a_7Tr(V_\mu V^\mu)Tr(TV_\nu)Tr(TV^\nu) + \frac{1}{4}a_{8}g^2 [Tr(T W_{\mu\nu})]^2 \nn \\
              &+& \frac{1}{2}ia_9Tr(TW_{\mu\nu})Tr(T[V^\mu,V^\nu]) + \frac{1}{2}a_{10}[Tr(T V_\mu)Tr(TV_\nu)]^2 \nn \\
              &+& a_{11}g\epsilon^{\mu\nu\rho\lambda}Tr(TV_\mu)Tr(V_\nu W_{\rho\lambda})
               \label{lag} \, .   
 \end{eqnarray} 
In (\ref{lag}), $V_\mu = (D_{\mu}U)U^\dagger$, $T = U\sigma^{3}U^\dagger$ and 
\be
D_\mu U = \partial_\mu U + i\frac{\sigma^k}{2}W^k_\mu U - ig'U\frac{\sigma^3}{2}B_\mu\, ,
\ee
with $W_{\mu\nu}=\sigma^kW^k_{\mu\nu}/2 = \partial_\mu W_\nu - \partial_\nu W_\mu + ig[W_\mu,W_\nu]$ is expressed in matrix notation.

This lagrangian, as any other effective theory, contains arbitrary coefficients, in this case called $a_i$, which have to be fixed by experiments or by matching the theory with a UV completion. 
The coefficients $a_2,a_3,a_9,a_{11}$ and $a_4,a_5,a_6,a_7,a_{10}$ contribute at tree level to the gauge boson scattering and represent anomalous triple and quartic gauge couplings respectively. They are not directly bounded by experiments.
On the other hand, the coefficients $a_0$, $a_1$ and $a_8$ in (\ref{lag}) are related to the electro-weak precision measurements parameters $S$, $T$ and $U$~\cite{peskin} and therefore
directly constrained by LEP precision measurements.\footnote{The authors of~\cite{barbieri} defined the complete set of EW  parameters which includes---in addition to  $S$, $T$ and $U$---$W$ and $Y$. The last two come from $O(p^6)$ terms and can be neglected in the present discussion.}

\subsubsection{Precision tests, custodial symmetry and the effective lagrangian}

The EW precision measurements test  processes in which  oblique corrections  play a dominant role with respect to the vertex corrections. This is why we can safely neglect the fermion sector (in our approximate treatment) and why the parameters 
$S$, $T$, $U$, $W$ and $Y$ represent such a  stringent phenomenological set of constraints for any new sector to be a candidate for EW symmetry breaking (EWSB). The good agreement between experiments and a single fundamental Higgs boson is encoded in the very small size of the above EW precision tests parameters. The idea of a fundamental Higgs boson is perhaps the most appealing  because of its extreme economy but it is not the only possibility  and what we  do here is to consider some strongly interacting new physics  whose role is providing masses for the gauge bosons in place of the Higgs boson. 

To express the precision tests constraints in terms of bounds for the coefficients of the 
low-energy lagrangian in \eq{lag} we have to take into account that the parameters $S$, $T$ and $U$ are defined as deviations from the SM predictions evaluated at a reference value for the Higgs and top quark masses. Since we are interested in substituting the SM Higgs sector, we keep separated the contribution to $S$ of the Higgs boson and write
\be
S_{H} + S = S_{EWSB} \, ,
\ee
and analog equations for $T$ and $U$. 
The contributions coming from the SM particles, including the GB, are not relevant because they appear on both sides of the equation. $S_H$ is given  by diagrams containing at least one SM Higgs boson propagator while $S_{EWSB}$ represents the contribution of the new symmetry breaking sector, except for contributions with GB loops only.
We thus find that, in the  chiral lagrangian (\ref{lag}) notation,
\bea
S_{EWSB} &=& -16\pi a_1\nn\\
\alpha_{em}T_{EWSB} &=& 2g^2a_0\nn\\
U_{EWSB} &=& -16\pi a_8 \label{a-STU}
\eea
The coefficients $a_0$, $a_1$ and $a_8$  typically have a scale dependence (and the same is true for $S_{H}$, $T_{H}$ and $U_{H}$) because they renormalize the UV divergences of the GB loops which yields a renormalization scale independent $S$, $T$ and $U$.  
One expects by dimensional analysis that $U\sim (m_Z^2/\Lambda^2)T \ll T$ and therefore $U$ is typically ignored. The relationships (\ref{a-STU}) have been used in \cite{bagger} to study the possible values of the effective lagrangian coefficients in the presence of SM Higgs boson with a mass larger than the EW precision measurements limits.

Using the results of the analysis presented in~\cite{barbieri}, taking as reference values  $m_H = 115$ GeV, $m_t = 178$ GeV and summing the 1-loop Higgs contributions, we obtain:
\bea
S_{EWSB} &=& -0.05 \pm 0.15 \nn \\ 
\alpha_{em} T_{EWSB} &=& 0.1 \pm 0.9 \label{S}
\eea
at the scale $\mu=m_Z$. We shall use these results  to set constraints to the coefficients of the effective lagrangian (\ref{lag}).

The smallness of the parameter $T$ can be understood as a consequence of the approximate $SU(2)_C$ custodial symmetry.
Due to this approximate symmetry we expect the couplings $a_{0,2,6,7,8,9,10,11}$ to be subdominant with respect to the custodial preserving ones.  Gauge boson scattering is then dominated by only two coefficients: $a_4$ and $a_5$. 

\subsubsection{Scattering amplitude}

Being interested in the EW symmetry breaking sector, we will mostly deal with longitudinally polarized vector bosons scattering because it is in these processes that the new physics plays a dominant role.
We can therefore make use of the equivalence theorem  (ET) wherein
the longitudinal $W$ bosons are replaced by the Goldstone bosons~\cite{ET}. This approximation is valid up to orders $m_W^2/s$, where $s$ is  the center of mass (CM) energy, and therefore---by also including the assumptions underlaying the effective lagrangian approach---we require our scattering amplitudes to exist in a range of energies such as $m_W^2\ll s\ll \Lambda^2$.

Assuming exact $SU(2)_C$, the  elastic scattering of gauge bosons  is described by a single amplitude $A(s,t,u)$. Up to $O(p^4)$, and by means of the lagrangian (\ref{lag}) we obtain~\cite{GL} 
\bea
A(s,t,u) & =& \frac{s}{v^{2}} 
 +\frac{4}{v^{4}}\left[ 2a_{5}(\mu )s^{2}+a_{4}(\mu )(t^{2}+u^{2})+\frac{{1}}{(4\pi )^{2}}\frac{{4 s^{2}+7\, (t^{2}+u^{2})}}{72}\right] \nn \\
 & -&\frac{1}{96\pi ^{2}v^{4}} \left[ t(s+2t)\log (\frac{-t}{\mu ^{2}})+u(s+2u)\log (\frac{-u}{\mu ^{2}})+3s^{2}\log (\frac{-s}{\mu ^{2}})\right]\label{amp} 
 \eea
where $s,t,u$ are the usual Mandelstam variables satisfying $s+t+u=0$ which in the CM frame and for any $1+2\rightarrow1'+2'$ process  can be expressed as a function of  $s$ and  the scattering angle $\theta$ as $t=-s(1-\cos\theta)/2$ and $u=-s(1+\cos\theta)/2$.

The GB carry an isospin $SU(2)_C$ charge $I=1$ and we can express any process in terms of isospin amplitudes $A_I(s,t,u)$ for $I=0,1,2$:
\begin{eqnarray}
A_{0}(s,t,u) & = & 3A(s,t,u)+A(t,s,u)+A(u,t,s)\nn \\
A_{1}(s,t,u) & = & A(t,s,u)-A(u,t,s)\nn \\
A_{2}(s,t,u) & = & A(t,s,u)+A(u,t,s) \label{isoamp} \, .
\end{eqnarray}

From the above results, we obtain the amplitudes for the scattering of the physical longitudinally polarized gauge bosons as follows:
\bea
A(W^{+}W^{-}\to W^{+}W^{-}) & = & \frac{1}{3}A_{0}+\frac{1}{2}A_{1}+\frac{1}{6}A_{2}\nn \\
A(W^{+}W^{-}\to ZZ) & = & \frac{1}{3}A_{0}-\frac{1}{3}A_{2} \nn
 \\
A(ZZ\to ZZ) & = & \frac{1}{3}A_{0}+\frac{2}{3}A_{2}\nn \\
A(WZ\to WZ) & = & \frac{1}{2}A_{1}+\frac{1}{2}A_{2}\nn \\
A(W^{\pm }W^{\pm }\to W^{\pm }W^{\pm }) & = & A_{2}\, .\label{Wamp}
\eea

It is useful to re-express the scattering amplitudes in terms of partial waves
 of definite angular momentum $J$
and isospin $I$ associated to the custodial $SU(2)_C$ group. 
These partial waves are denoted $t_{IJ}$ and are defined,
in terms of the  amplitude $A_I$ of (\ref{isoamp}), as
\be
  t_{IJ}=\frac{1}{64\,\pi}\int_{-1}^1\,d(\cos\theta)
\,P_J(\cos\theta)\,A_I(s,t,u)\;.
\ee
Explicitly we find:
\begin{eqnarray}
 t^{(2)}_{00}&=&\frac{s}{16\,\pi v^2}, \quad
 t^{(4)}_{00}=\frac{s^2}{64\,\pi v^4}
\left[\frac{16(11a_5+7a_4)}{3}
+\frac{17/3-50 \log(s/\mu^2)/9+4\, i\,\pi}{16\,\pi^2}
\right],\nonumber\\
 t^{(2)}_{11}&=&\frac{s}{96\,\pi v^2},  \quad
 t^{(4)}_{11}=\frac{s^2}{96\,\pi v^4}\left[4(a_4-2a_5)
+\frac{1}{16\,\pi^2}\left(\frac{1}{9}+\frac{i\,\pi}{6}\right)
\right],\nonumber\\
 t^{(2)}_{20}&=&\frac{-s}{32\,\pi v^2}, \quad
 t^{(4)}_{20}=\frac{s^2}{64\,\pi v^4}\left[\frac{32(a_5+2a_4)}{3} 
+\frac{17/6-20\log(s/\mu^2)/9
+i\,\pi}{16\,\pi^2}
\right] \, ,
\label{pertamplis}
\end{eqnarray}
where the superscript refers to the corresponding power of momenta.

The contributions from $J\geq2$ start at order $p^4$ and turn out to be irrelevant for our purpose.
The $I=1$ channel is related to an odd spin field due to the Pauli exclusion principle. 
The $(I=2, J=0)$ channel has a dominant minus sign which, from a semiclassical perspective, indicates that this channel is repulsive and we do not expect any resonance with these quantum numbers. 

The amplitudes (\ref{amp}) (or, equivalently (\ref{pertamplis})) show that, for $s\gg m_W^2$, the elastic scattering of two longitudinal polarized gauge bosons is observed with a probability that increases with the CM energy $s$. We  expect that for sufficiently large energies the quantum mechanical interpretation of the $S$-matrix will be lost. We explicitly show the unitarity bound thus obtained as a dashed line in the plots presented below in Figures (\ref{fig3}) and (\ref{fig4}) at the end of the paper.

The effective lagrangian (\ref{lag}) and gauge boson scattering were extensively discussed  in~\cite{ww1}.

\subsection{Limits and constraints}

If we knew all the coefficients of the lagrangian (\ref{lag}), and $a_4$ and $a_5$ in particular, we could fully predict gauge boson scattering at the LHC.  We therefore turn now to  what is known about them in order to review all current constraints on their possible values and compare them with   the limits  which are going to be explored given the expected LHC sensitivity. As we shall see,  these two crucial coefficients are poorly  known quantities which furthermore will not be fully explored at the LHC.

\subsubsection{LHC sensitivity}

First of all, let us consider the capability of the LHC of exploring the coefficients $a_4$ and $a_5$ of the effective lagrangian (\ref{lag}). This has been discussed most recently  in~\cite{eboli} by comparing cross sections with and without the operator controlled by the corresponding coefficient. They consider scattering of $W^+W^-$,  $W^\pm Z$ and $ZZ$ ($W^\pm W^\pm$ gives somewhat weaker bounds) and report limits (at 99\% CL) that we take here to be
\bea
-7.7 \times 10^{-3} \leq a_4 \leq 15 \times 10^{-3} \nn\\
-12 \times 10^{-3} \leq a_5 \leq 10 \times 10^{-3} \, . \label{limLHC}
\eea
The above limits are obtained considering as non-vanishing only one coefficient at the time.  It is also possible to include both coefficients together and obtain a combined (and slightly smaller) limit. We want to be conservative and therefore use (\ref{limLHC}).  Comparable limits were previously found in the papers of ref.~\cite{pre-eboli1}.
 
To put these results in perspective, limits roughly one order of magnitude better can be achieved by a linear collider~\cite{LC}.

\subsubsection{EW precision measurements: indirect bounds}

Bounds on the coefficients $a_4$ and $a_5$ can be obtained by including their effect (at the one-loop level) into low-energy and $Z$ physics precision measurements. We call them indirect bounds since they only come in at the loop level.

As expected, these bounds turn out to be rather weak~\cite{eboli} :
\bea
-320 \times 10^{-3} \leq a_4 \leq 85 \times 10^{-3} \nn \\
-810 \times 10^{-3} \leq a_5 \leq 210 \times 10^{-3} \,  \label{indirect}
\eea 
at 99\% C.L. and for $\Lambda=2$ TeV. Comparable bounds were previously found in the papers in ref.~\cite{pre-eboli2}. As before, slightly stronger bounds can be found by a combined analysis.

Notice that  the $SU(2)_C$ preserving triple gauge coupling $a_3$ has not been considered in the  computations leading to the previous limits. Once its contribution is taken into account, the LHC sensitivity and the indirect bounds presented here are  slightly modified although the ranges shown are not  changed drastically. 

\begin{figure}[hbtp]
\begin{center}
\includegraphics[width=5in]{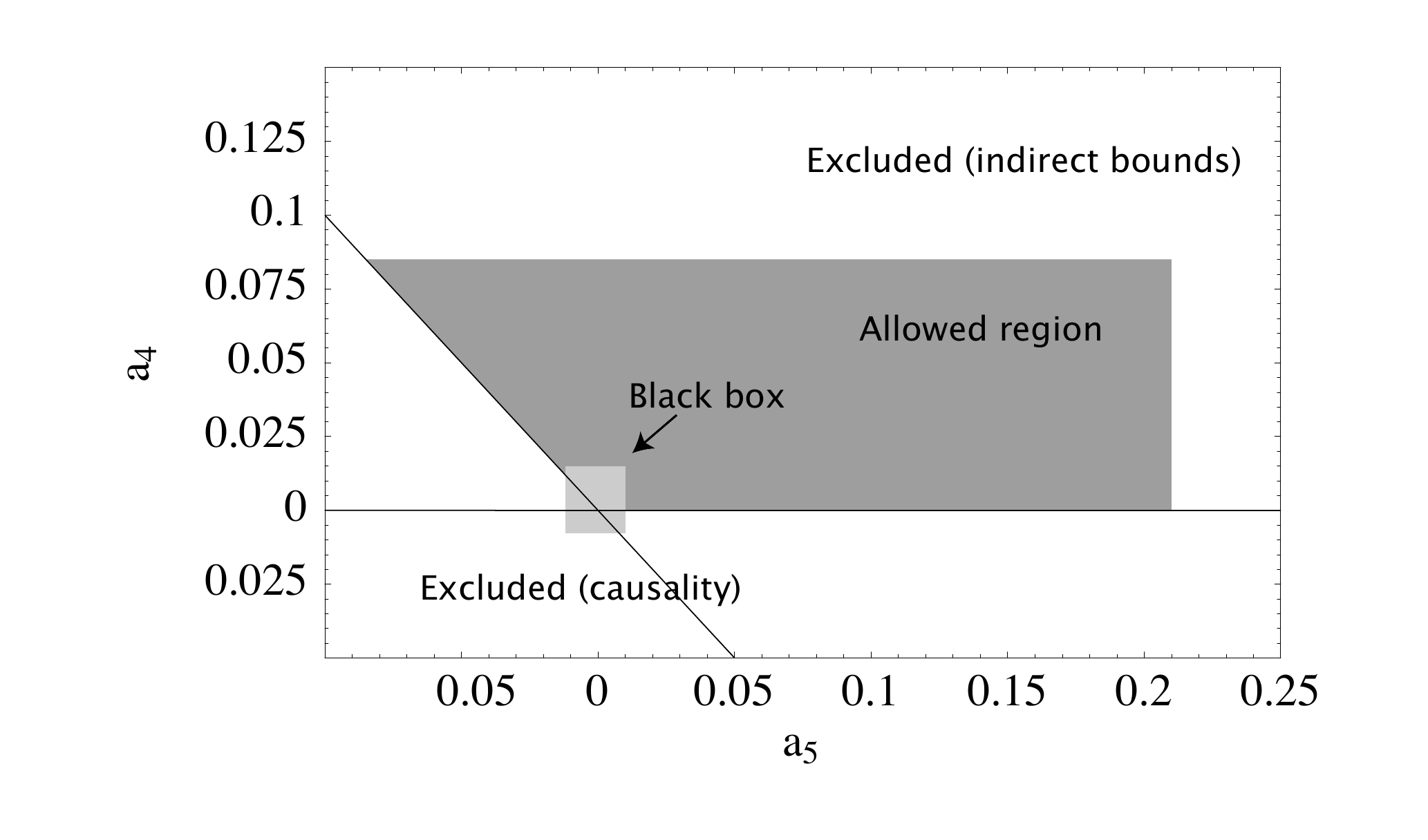}
\caption{\small  The region of allowed values in the $a_4$-$a_5$ plane (in gray) as provided by combining indirect bounds and causality constraints. Also depicted, the region below which LHC will not be able to resolve the coefficients (Black box). \label{fig1}}
\end{center}
\end{figure}

\subsubsection{Unitarity, analyticity and causality}

The requirement of  unitary of the theory forces the cut off of the lagrangian (\ref{lag}) to be $\Lambda\leq1.3$ TeV but does not impose any constraint on the coefficients $a_{i}$. Other fundamental assumptions like causality and analyticity of the $S$-matrix do give rise to interesting constraints.

In particular, the causal and analytic structure of the amplitudes leads to bounds on the possible values the two coefficients $a_4$ and $a_5$ can assume. This is well known in the context of chiral lagrangians for the strong interactions~\cite{pham} and can be extended with some caution 
to the weak interactions.
It can be shown in fact that the second derivative with respect to the center of mass energy of the forward elastic scattering amplitude of two GB is bounded from below by a positive integral of the total cross section for the transition $2 \pi \rightarrow everything$. The coefficients $a_4$ and $a_5$ enter this amplitude and one can use the mentioned result to bound them. 

The most stringent bounds  come from  the requirement that the underlying theory respects causality.
For a discussion on this point and the  different result found in ref.~\cite{Distler} see~\cite{vecchi}.
The causality bound can be understood by noticing that, given a classical solution of the equations of motion, one can study the classical oscillations around this background interpreting the motion of the quanta as a scattering process on a macroscopic object~\cite{Rattazzi}. If the background has a constant gradient, the presence of superluminal propagations sum up and can in principle become manifest in the low-energy regime. 

Following the argument in~\cite{Rattazzi}, by means of  solutions of the type $\pi_{0}=\sigma^iC_\mu x^\mu$ we obtain the free equations of motion for the oscillations around this background. All three of them can be written as
\be 
p^2\left( 1+O(a)\right) + \frac{a}{v^4} \left( C\cdot p\right)^2 = 0\, ,  
\ee
with $a=a_{4}$ or $a=a_4+a_5$. In this derivation we made use of the assumption $C^2\ll \Lambda^4$ which is necessary to ensure a perturbative expansion in the framework of the effective theory. 
The above relations represent a subluminal group velocity only in the case $a\geq0$. 
These classical results can be implemented in a quantum framework provided we take into account that all of the coefficients $a_i$ are formally evaluated at a scale $\mu< \Lambda$ through a matching procedure between the UV theory and the lagrangian (\ref{lag}). 

In conclusion, the causality constraints can  be taken to be~\cite{vecchi}
\bea
a_4(\mu) &\geq& \frac{1}{12} \frac{1}{(4\pi)^2} \log \frac{\Lambda^2}{\mu^2}\nn \\
a_4(\mu) + a_5(\mu) &\geq& \frac{1}{8} \frac{1}{(4\pi)^2} \log \frac{\Lambda^2}{\mu^2} \, . 
\label{unitbound}
\eea

Notice that the constraints in \eq{unitbound} remove a quite sizable region (most of the negative values, in fact) of values of the parameters $a_4$ and $a_5$ allowed by the indirect bounds alone.  Fig.~\ref{fig1} summarizes the allowed values in the $a_4$-$a_5$ plane and compare it with LHC sensitivity.

\subsection{EW precision measurements: direct (model dependent) bounds}
\label{sec:db}

Given the results in Fig.~\ref{fig1}, we can ask ourselves how likely are the different values for the two coefficients $a_4$ and $a_5$ among those within the allowed region. Without further assumptions, they are all equally possible and no definite prediction is possible about what we are going to  see at the LHC. 

In order to gain further information, we would like to find  relationships between these two coefficients and between them and those of which the experimental bounds are known.To accomplish this, we have to introduce some more specific assumptions about the ultraviolet (UV) physics beyond the cut off of the effective lagrangian.  We do it in the spirit of using as much as we know in order to guess what is most likely to be found. 

A  step in this direction consists in assuming a specific UV completion beyond the cut off of the effective lagrangian in \eq{lag}. The two most likely scenarios which can be studied with the effective lagrangian approach are a confining theory (essentially the gauge sector of a strongly interacting model of a rescaled QCD) and the strongly coupled regime of a  model like the SM Higgs sector in which the Higgs boson is heavier than the cut off. For each of these two scenarios it is possible to derive more restrictive relationships among the coefficients of the EW lagrangian and in particular we can relate parameters like $a_0$ and $a_1$ to $a_4$ and $a_5$. These new relationships make possible to use EW precision measurements to constrain the possible values of the coefficients $a_4$ and $a_5$.

\subsubsection{Large-$N$ scenario}

This scenario is based on a new $SU(N)$ gauge theory coupled to new fermions charged under the fundamental representation. By analogy with QCD these particles  are invariant under a flavor chiral symmetry containing the gauged $SU(2)_L\times U(1)_Y$ as a subgroup. 
Let us consider the case in which no other GB except the 3 unphysical ones are present and therefore the chiral group has to be $SU(2)_L \times SU(2)_R$, with $U(1)_Y\subset SU(2)_R$. The new strong dynamics  leads directly to  EWSB through the breaking of the axial current conservation under the confining scale around $4 \pi v$ and to the appearance of an unbroken $SU(2)_{L+R}=SU(2)_C$ custodial symmetry. Following these assumptions, there are no bounds on the new sector from the parameter $T$  and the relevant constraints come from the $S$ parameter only.\footnote{We are not concerned here with the fermion masses and therefore we can bypass most of the problems plaguing technicolor models.}

At energies under the confining scale, the strong dynamics can be described in terms of the hadronic states. Their behavior can be simplified by making use of the large-$N$ approximation. The main result is that the resonances appearing as low-energy degrees of freedom have negligible self-interactions with respect to the couplings to the GB. This limit turns out to be a good approximation of  low-energy QCD even if $N$ is not large.

The large-$N$ approximation allows us to readily estimate the coefficients of the effective lagrangian. The coefficients $a_i$ are scale independent at the leading order in the $1/N$ expansion. At this order, we find that $a_4$ and $a_5$ are finite and (by transforming the result of~\cite{N} for QCD)
\be
a_4 = - 2 a_5 = - \frac{1}{2} a_1 \, ,
\ee
which provide us with the link between gauge boson scattering and EW precision measurements---the coefficient $a_1$ being directly related to the parameter $S$ as indicated in \eq{a-STU}.

In a more refined approach, the  non-perturbative effects have been integrated out giving rise to a constituent fermion mass and a gauge condensate. The chiral lagrangian is a consequence of the integration of these massive states. The result becomes~\cite{bijnens}:
\bea
a_{4} &=& \frac{N}{12(4\pi)^2} \nn \\
a_{5} &=& -\left(\frac{1}{2} + \frac{6}{5}\langle G^2\rangle \right)a_{4}\, , \label{QCD1}
\eea
where $\langle G^2\rangle$ is an average over gauge field  fluctuations. The latter is a positive and order 1 free parameter that encodes the dominant soft gauge condensate contribution  which there is no reason to consider as a negligible quantity. Without these corrections the result is equivalent to those obtained considering the effect of a heavier fourth family. Causality requires $\frac{6}{5}\langle G^2\rangle \leq \frac{1}{2}$ and therefore we will consider values of $\langle G^2\rangle$ ranging between $0<\langle G^2\rangle<0.5$.

The coefficients $a_i$ are scale independent at the leading order in the $1/N$ expansion.

The $S$ parameter gives stringent constraints on $N$:
\be
S_{EWSB} = \frac{N}{6\pi}\left(1+\frac{6}{5}\langle G^2\rangle\right) 
\ee
which is slightly increased by the strong dynamics with respect to the perturbative estimate, in good agreement with the non-perturbative analysis given in~\cite{peskin}. From the bounds on $S_{EWSB}$, we have $N$ $<$ 4 ($2\sigma$) and N $<$ 7 ($3\sigma$) respectively. 

The relevant  bounds on $a_4$ is then obtained via $a_1$ and yields
\be
0<a_4 < \frac{S_{EWSB}}{32\pi}\, . \label{QCD2}
\ee
We are going to use the bounds given in \eq{QCD1} and \eq{QCD2}. 

Taking  $a_1$ at the central value of $S_{EWSB}$ gives $a_4<0$, which is  outside the causality bounds. This is just a reformulation in the language of effective lagrangians of the known disagreement with EW precision measurements of most models of strongly interacting EW symmetry breaking.

We expect vector and scalar resonances to be the lightest states. The high spin or high $SU(2)_C$ representations considered earlier are typically bound states of more than two fermions and therefore more energetic. Their large masses make their contribution to the $a_i$ coefficients subdominant. 

The relations (\ref{unitbound}) and (\ref{QCD1}) satisfied by the model imply that $-a_4<a_5<-a_4/2$, an indication that  scalar resonances give contributions comparable with the vectorial ones in the large-$N$ limit. If vectors had been the only relevant states, the relation would have been $a_4=-a_5$. 

It is useful to pause and  compare this result with that in low-energy QCD. 

Whereas in the EW case we find that the large-$N$ result indicates the importance of having low-mass scalar states, the chiral lagrangian of low-energy QCD has the corresponding parameters  $L_1$ and $L_2$ saturated by the vector states alone. This vector meson dominance is supported by the experimental data and  in agreement with the large-$N$ analysis, which in  the case of the group $SU(3)$ is different from that of the EW group $SU(2)\times U(1)$.

Even though the scalars have little  impact on the effective lagrangian parameters of low-energy QCD, they turn out to be relevant  to fit the data at energies larger than the $\rho$ mass where the very wide $\sigma$ resonance appearing in the amplitudes is necessary~\cite{sannino}.
One may ask if something similar applies to the EWSB sector, it being described by a similar low-energy action. This can be  seen by looking at the contribution of a single vector to the tree-level fundamental amplitude:
\be
A(s,t,u) = \frac{s}{v^2} - \frac{3M_V^2 s}{\hat{g}^2 v^4} + \frac{M_V^4}{\hat{g}^2 v^4}\left(\frac{u - s}{t - M_V^2} + \frac{t - s}{u - M_V^2} \right) 
\ee
with $\hat{g}$ (not to be interpreted as a gauge coupling) and $M_V^2$ representing the only two parameters entering up to order $p^4$.
The limit $s \ll M_V^2$ corresponds to integrate the vector out and gives the low energy theorem with the previously mentioned $a_{4} = -a_{5} = 1/(4\hat{g}^2)$, while the opposite limit $s\gg M_V^2$ is not well defined. 
The condition $M_V^2=\hat{g}^2v^2/3$ erases the linear term but cannot modify the divergent behavior of the forward and backward scattering channels. In fact we still find the asymptotic form $t_{00}(s) \simeq \hat{g}^2/(36\pi)\log(s/M_V^2)$ which has to be roughly less than one half to preserve unitarity.
This shows why models with only vector resonances cannot move the UV cut off too far from the vector masses, as opposed  to what happens in the case of scalar particles.

\begin{figure}[hbtp]
\begin{center}
\includegraphics[width=5.5in]{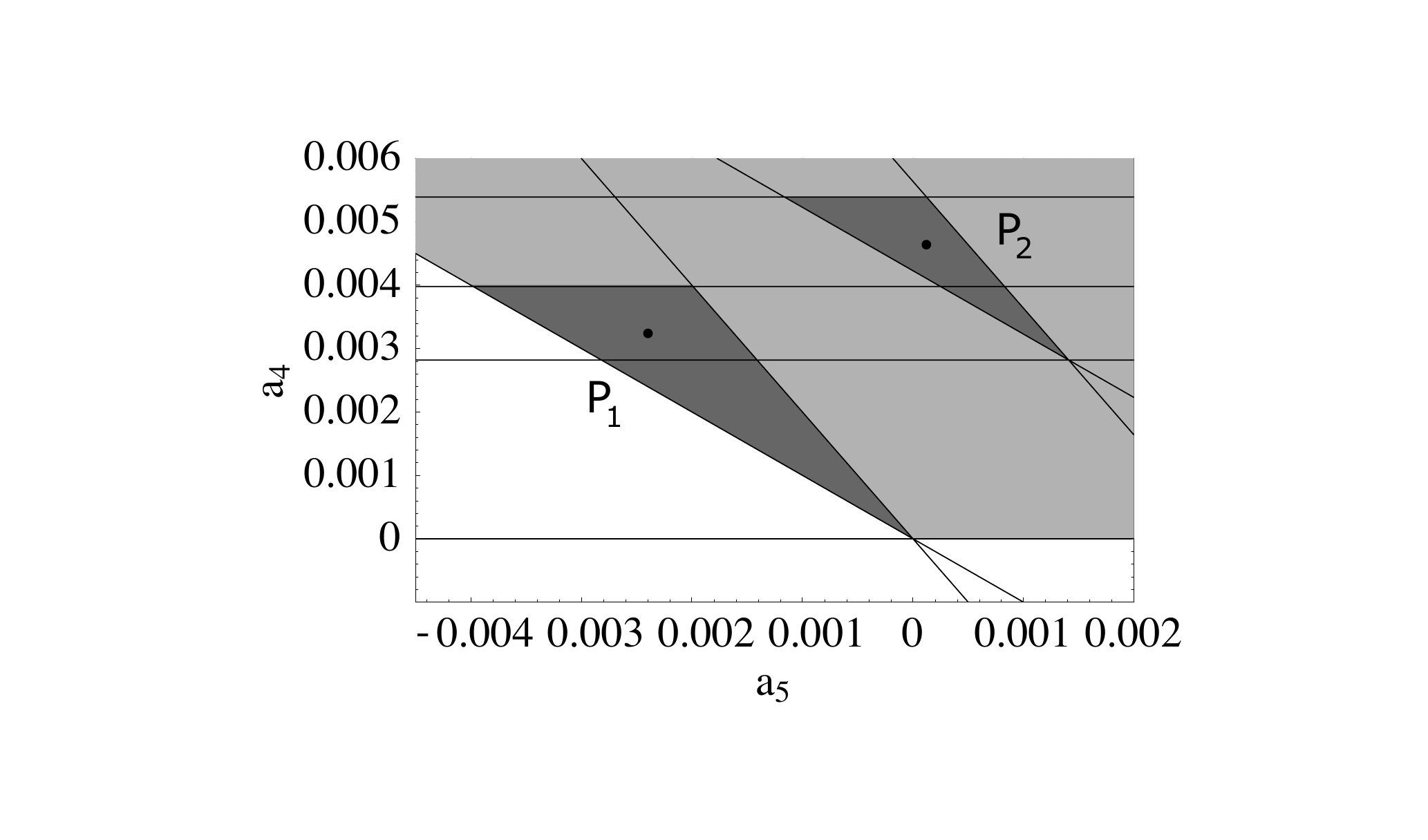}
\caption{\small  Model-dependent bounds for the coefficients for the large-$N$ scenario. All points within the gray background are allowed by the causality bounds. The two upper horizontal lines mark the bounds from EW precision tests. The bigger dark triangle contains the allowed values when neglecting the running of the coefficients. The smaller includes the first cutoff-dependent subleading correction in the large $N$ expansion. See the text for a discussion of this point. 
Two representative points are indicated: $P_1$ and $P_2$.  Notice that the range of this figure is all within the black box of Fig.~\ref{fig1}. \label{fig2b}}
\end{center}
\end{figure}

The larger dark triangle in Fig.~\ref{fig2b} shows  the allowed values for the coefficients $a_4$ and $a_5$ as given by  \eq{QCD1} and \eq{QCD2}. The gray background is drawn according to the causality constrain which is assumed  scale independent  to be consistent with the leading large-$N$ result.

As an estimate of the first subleading corrections we can include the  running of the coefficients  and obtain rescaled causality bounds and a shrinkage of the region defined by \eq{QCD2} due to the different anomalous dimensions of the chiral coefficients. This is represented in Fig.~\ref{fig2b} by the second and smaller dark triangle where a cutoff $\Lambda = 1.3$ TeV has been chosen.
The two dark triangles thus obtained give an idea of the uncertainty of the scenario and all values of the coefficients $a_4$ and $a_5$  in and between these two regions are equally acceptable.

\begin{figure}[hbtp]
\begin{center}
\includegraphics[width=5.5in]{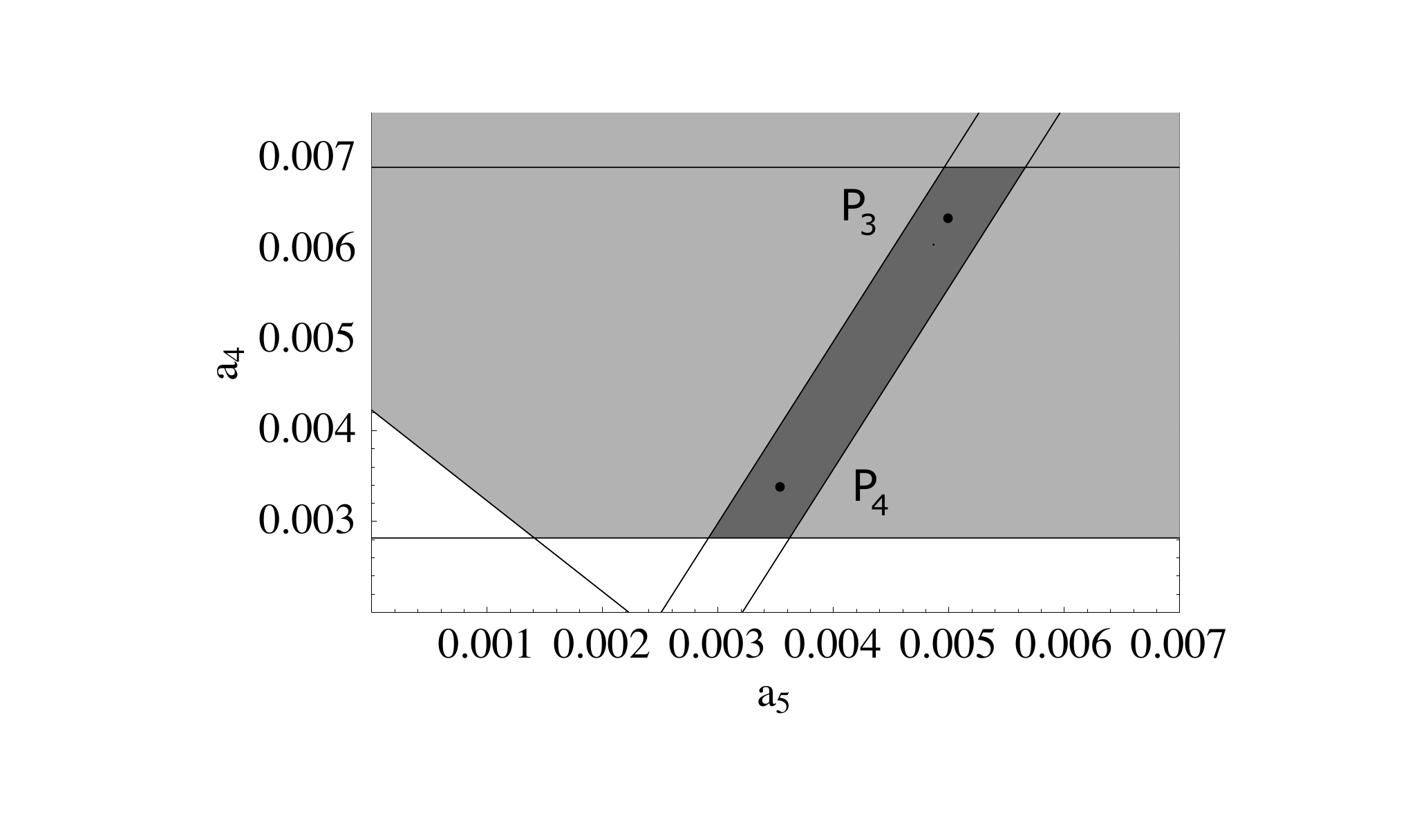}
\caption{\small  Model-dependent bounds for the coefficients in the heavy-Higgs scenario. All points within the gray background are allowed by the causality bounds. The upper horizontal line marks the bound from EW precision tests. The two diagonal lines correspond to the two values of the Higgs masses: 2 and 2.5 TeV. Two representative points are indicated: $P_3$ and $P_4$. Notice that the range of this figure is all within the black box of Fig.~\ref{fig1}. \label{fig2a}}
\end{center}
\end{figure}

\subsubsection{Heavy-Higgs scenario}

This scenario is a bit more contrived than the previous one and a few preliminary words are in order.

A scalar Higgs-like particle violates unitarity for masses of the order of 1200 GeV~\cite{willenbrock}. Moreover, the mass of the Higgs is proportional to its self coupling and from a naive estimate we expect the perturbation theory to break down at $\lambda \sim 4 \pi$, that is $m_H\sim1300$ GeV. 
What actually happens in the case of a non-perturbative coupling is not known. Problems connected with triviality are not rigorous in non-perturbative theories and therefore the hypothesis of a heavy Higgs cannot be ruled out by this  argument.  

As long as we intend such a heavy Higgs boson only as a modeling of the UV completion of the EW effective lagrangian, we can   study this scenario by assuming a Higgs mass between 2 and 2.5 TeV. Even though we cannot expect the perturbative calculations to be  reliable at these scales, they may still  provide some insight  into   the strongly interacting behavior.

The effective lagrangian parameters in the case of a heavy Higgs can be computed by retaining only the leading logarithmic terms to yield: 
\be
a_4 = - a_1 \quad \mbox{and} \quad a_4 = 2 a_5 \, ,
\ee 
which contains the link between gauge boson scattering and the coefficient $a_1$ we need.
A more complete computation~\cite{herrero} gives
\bea
a_4(m_Z)&=&-\frac{1}{12}\frac{1}{(4\pi)^2}\left( \frac{17}{6}-\log\frac{m_H^2}{m_Z^2}\right)\nn \\ 
a_5(m_Z)&=&\frac{v^2}{8m_H^2}-\frac{1}{24}\frac{1}{(4\pi)^2}\left( \frac{79}{3}-\frac{27\pi}{2\sqrt{3}}-\log\frac{m_H^2}{m_Z^2}\right)
\eea
and
\be
S_{EWSB}=\frac{1}{12\pi}\left( \log\frac{m_H^2}{m_Z^2}-\frac{5}{6}\right) \, .
\ee

The causality constrain (\ref{unitbound}) applied to the above equations implies a bound on the possible values of the cutoff $\Lambda$ compared to $m_H$. An effective lagrangian cutoff consistent with LHC physics yields a Higgs mass at least of the order of 2 TeV. 

Putting these equations together,  we obtain:
\bea
a_4 = \frac{1}{16\pi}\left( S_{EWSB}-\frac{1}{6\pi}\right) \nn \\
 a_4 = 2a_5 - \frac{v^2}{4m_H^2} + \frac{1}{12}\frac{1}{(4\pi)^2}\left(\frac{141}{6}-\frac{27\pi}{2\sqrt{3}}\right) 
\eea
As before in the large-$N$ scenario, the central value of $S_{EWSB}$ yields a value of $a_4$ outside the causality bounds.

 In this scenario there  appears a very small non-vanishing $T$ parameter from loop effects which however gives no relevant bounds to the $a_{4}$ and $a_{5}$ couplings because
\be
\alpha_{em}T_{EWSB} = -g'^2\frac{9}{16\pi}S_{EWSB} \ll S_{EWSB} \, .
\ee

At this point we can collect these results with those of the previous section and conclude that in both scenarios under study, the limits on the coefficients $a_4$ and $a_5$ are well below LHC sensitivity (compare Fig.~\ref{fig1} and Fig.~\ref{fig2b} and \ref{fig2a}). If this is the case,
the  LHC will probably not be able to resolve the value of these coefficients because they are too small to be seen. It goes without saying that this can only be a provisional conclusion in as much as in many models the relations among the coefficients we utilize  can be made weaker by a variety of modifications which make the models more sophisticated.  Accordingly, our bounds will not apply and the LHC may indeed measure $a_4$ or $a_5$ and we will then know that the UV physics is not described by the simple models we have considered.

\subsubsection{A comment about Higgsless models}

Higgsless models~\cite{higgsless} have been proposed to solve the hierarchy problem. They describe a gauge theory in a 5D space-time that produces the usual tower of massive vectors on the 4 dimensional brane (our world). The lightest Kaluza-Klein modes are  interpreted as the $W^{\pm}$ and $Z^0$ while those starting at a mass scale $\Lambda$, represent a new weakly coupled sector. 

The scale of unitarity violation is automatically raised to energies larger than 1.3 TeV because the term in the amplitude  linearly increasing with the CM energy $s$ is not present in these models.
Every 5D model, whatever the curvature, has this property and  fine tuning is neither required nor possible. 
For this reason,  a saturation of the unitarity bound of the term of the amplitude linear in $s$ with just a few vector states, as done in \cite{perelstein}, cannot be considered a characteristic signature of the Higgsless models.

These 5D models fear no better than technicolor when confronted by EW precision measurements. There exists an order 1 mixing among the heavy vectors which contribute a tree level $W_\mu^3-B_\nu$ exchange and consequently a $S_{EWSB}\propto1/(gg')$. 
In 5D notation and for the simplest case of a flat metric, $S_{EWSB} = O(1)/g^2\simeq R/g^2_{(5)}$, in agreement with~\cite{BPR}.
This result can be ameliorated by the introduction of a warped 5D geometry, or boundary terms or even by a de-localization of the matter fields~\cite{cacciapaglia}.
In a certain sense these fine tuning can be seen as a 5D analog of the walking effect on a QCD-like Technicolor.

As it will become clear in the next section, our general analysis of the resonant spectrum relies on the presence of the linear term in $s$ and therefore any 5D Higgsless model is a priori excluded. Nevertheless, since we already know what is the spectrum, we can give some indicative result of what an Higgsless model implies for the coefficients $a_4$ and $a_5$. 

These models present the relation $a_4=-a_5$ which is characteristic of all models with vector resonances only. 
This line in the $a_4$-$a_5$ plane of Fig.~\ref{fig2b} lies on the causality bound and coincides with the large-$N$ scenario in which the strong dynamical effect $\langle G^2\rangle$ is maximal or, equivalently, in the case in which the scalar resonances are excluded. If we content ourselves with an estimate in the 5D flat space approximation we can write some explicit result~\cite{simmons}.
For example, the asymptotic behavior of $t_{00}$ in the case of a flat 5D geometry is found to be 
\bea
t_{00}\sim\frac{M_1^2}{\pi^3v^2}\log\left(\frac{s}{M_1^2} \right) 
\label{higgslessunit}
\eea
and represents an upper bound on the mass $M_1$ of the lightest massive excitation of the $W^\pm,Z^0$.

The coefficients $a_4$ is related via deconstruction to $a_1$.  We find that
\be
a_4 = -\frac{1}{10} a_1 \, , 
\ee
and therefore, 
\be
a_4=-a_5 = \frac{\pi^2}{120}\frac{v^2}{M_1^2}=\frac{S_{EWSB}}{160 \pi}\, .
\ee
The constraints on $S$ of \eq{S}  lead to $M_1>$ 2.5 TeV which implies a violation of unitarity, and consequently the need of a UV completion for the 5D theory, at the scale $\sim M_1^2$. 

The parameters $a_4$ and $a_5$ are---as in the other scenarios considered---too small to be directly detected at the LHC. The large mass $M_1$ of the first vector state makes it hard for the LHC to find it. 

In case of a warped fifth dimension these relations are slightly changed but the tension existing between the unitarity bound~(\ref{higgslessunit}) (which requires a small $M_1^2$ to raise the cut off above 1.3 TeV) and the $S$ parameter (which requires a large $M_1^2$) remains a characteristic feature of models with vector resonances only.

\vskip1.5em
\section{Experimental signatures: resonances} 
\label{sec:res}

Even though the values of the coefficients may be too small for the LHC, the unitarity of the amplitudes is going to be violated at a scale around 1.3 TeV unless higher order contributions are included. Following the well-established tradition of unitarization in the strong interactions, we  consider the Pad\'e approximation, also known as the inverse amplitude method (IAM)~\cite{pade}. Other unitarization procedure have been used in the literature but we find them less compelling than IAM because they introduce further (unknown) parameters.

This procedure is carried out in the language of the partial waves introduced in (\ref{pertamplis}). In fact, using analytical arguments we  find that
\begin{equation}
t_{IJ}(s)=\frac{t^{(2)}_{IJ}}{1-t_{IJ}^{(4)}/t^{(2)}_{IJ}}+ O(s^3) \, .
\label{IAM}
\end{equation}
Equation (\ref{IAM}) is the IAM, which
 has given remarkable results describing meson interactions, 
having a symmetry breaking pattern almost identical to our present case. 
Note that this amplitude respects strict elastic unitarity, while keeping the
correct  low energy expansion. Furthermore, the extension of
(\ref{IAM}) to the complex plane can be justified using
dispersion theory. 
In particular, it
has the proper analytical structure and, eventually,
poles in the second Riemann sheet 
for certain $a_4$ and $a_5$ values, that can be interpreted as resonances.
Thus, IAM formalism can 
describe resonances without increasing 
the number of parameters and respecting chiral symmetry
and unitarity. 

By inspection of \eq{IAM}, the IAM yields the following masses and widths of the first resonances:
\be
\label{scalar0}
m_{S}^{2}=\frac{{4v^{2}}}{\frac{{16}}{3}\left[11a_{5}(\mu )+7a_{4}(\mu )\right]+\frac{{1}}{16\pi ^{2}}\left[ \frac{{51-50\log (m_{S}^{2}/\mu ^{2})}}{9}\right] } \, , \quad
\Gamma _{S}=\frac{{m_{S}^{3}}}{16\pi v^{2}}\, ,
\ee
for scalar resonances, and
\be
\label{vector1}
m_{V}^{2}=\frac{{v^{2}}}{4\left[a_{4}(\mu )-2a_{5}(\mu )\right]+\frac{{1}}{16\pi ^{2}}\frac{{1}}{9}} \, ,
\quad
\Gamma _{V}=\frac{{m_{V}^{3}}}{96\pi v^{2}} \, ,
\ee
for vector resonances.

A few words of caution about the IAM approach are in order.

The resonances thus obtained represent the lightest massive states we encounter (above the $Z$ pole) in each channel which are necessary in order for the amplitude to respect unitarity. These resonances are not  the only massive states produced by  the  non-perturbative sector but those with  higher masses give a contribution that is subdominant with respect to the IAM prediction and can safely be ignored.

Since we neglect $O(s^3)$ terms, the regime $s\sim m_{res}^2$ is not completely trustable. The larger the resonance peak, the larger the error and therefore we expect the IAM prediction to give  good results only in the case of very sharp resonances. This is the reason behind the  success of the IAM  for the  vector resonances in QCD  as opposed to  the more problematic very broad scalar $\sigma$. 

Similarly, if we integrate a Higgs boson at the tree level and substitute the $a_{4}$ and $a_5$ parameters we find in the IAM formula, we obtain a value for the scalar resonance mass given by \eq{scalar0} which is smaller, that is $m_S = 3m_H/4$. 

Nevertheless, we consider the IAM result a remarkable prediction, given the very small amount of information needed. 

One way to check the  reliability of this method 
consists in separating the $a_{4,5}$ plane  into three areas depending on the predicted lowest laying resonances being a vector, a scalar or even both of them. This partition follows the coefficients patterns one expects by studying the tree level values for $a_4$ and $a_5$.

Another check on the consistency of the method 
is obtained by taking the unrealistic example in which $a_{4}=a_{5}=0$. In this case one finds a pole at an energy $s >(4\pi v)^2$---at which we already know unitarity is violated---thus indicating the unreliability of the input. More generally, a naive estimate---based on integrating out massive states like in the vector meson dominance of QCD---shows that for resonance masses $M$ between the range of hundreds GeV and a few TeV we should expect $a\simeq v^2/M^2$ from $10^{-2}$ to $10^{-3}$ which agrees with the IAM formula. 

Gauge boson scattering and the presence of resonances have previously been discussed in a number of papers~\cite{ww2,dobado}.

\subsection{Parton-level cross sections}

Our plan is to choose two representative points for each of the considered scenarios  in the allowed $a_4$-$a_5$ region  and then find the first resonances appearing in the $W_LW_L$ elastic scattering using the IAM approximations. The  points are shown in Fig.~\ref{fig2b} and \ref{fig2a}. We take
\be
P_1:\; \left\{ \begin{array}{ccc}  a_4 &=&  3.5  \times 10^{-3} \\
 a_5 &=& -2.5  \times 10^{-3} \end{array}
 \right.
 \quad \mbox{and} \quad 
P_2:\; \left\{ \begin{array}{ccc}  a_4 &=&  4.5  \times 10^{-3} \\
 a_5 &=& 0.2  \times 10^{-3} \end{array}
 \right.
\ee
for the large-$N$ scenario and
\be
P_3:\;\left\{ \begin{array}{ccc}  a_4 &=& 6.5  \times 10^{-3} \\
 a_5 &=& 5.0  \times 10^{-3} \end{array}
 \right.
 \quad \mbox{and} \quad 
P_4:\; \left\{ \begin{array}{ccc}  a_4 &=&  3.5  \times 10^{-3} \\
 a_5 &=& 3.5  \times 10^{-3} \end{array}
 \right.
\ee
for the heavy-Higgs scenario. 

The first pair corresponds to having  vector resonances at
\be
\left\{ \begin{array}{ccc}  m_V &=&  1340\; \mbox{GeV} \\
 \Gamma_V &=&  128\; \mbox{GeV}\end{array}
 \right.
 \quad \mbox{and} \quad 
 \left\{ \begin{array}{ccc}  m_V &=&  1900\; \mbox{GeV} \\
 \Gamma_V &=& 370\; \mbox{GeV} \end{array}
 \right.
\ee
together with   very broad scalar states,
while the second pair to  scalar resonances at
\be
\left\{ \begin{array}{ccc}  m_S &=&  660\; \mbox{GeV} \\
 \Gamma_S &=&  92\; \mbox{GeV}\end{array}
 \right.
 \quad \mbox{and} \quad 
 \left\{ \begin{array}{ccc}  m_S &=&  820\; \mbox{GeV} \\
 \Gamma_S &=& 175\; \mbox{GeV} \end{array}
 \right.
\ee
These points are representative of the possible values and span the allowed region.
The resonances become heavier, and therefore less visible at the LHC, for smaller values of the coefficients. Accordingly, whereas points $P_1$ and $P_3$ give what we may call an ideal scenario,  the other two show a situation that will be difficult to discriminate at the LHC.

\begin{figure}[h]
\hbox{\includegraphics[width=3.5in]{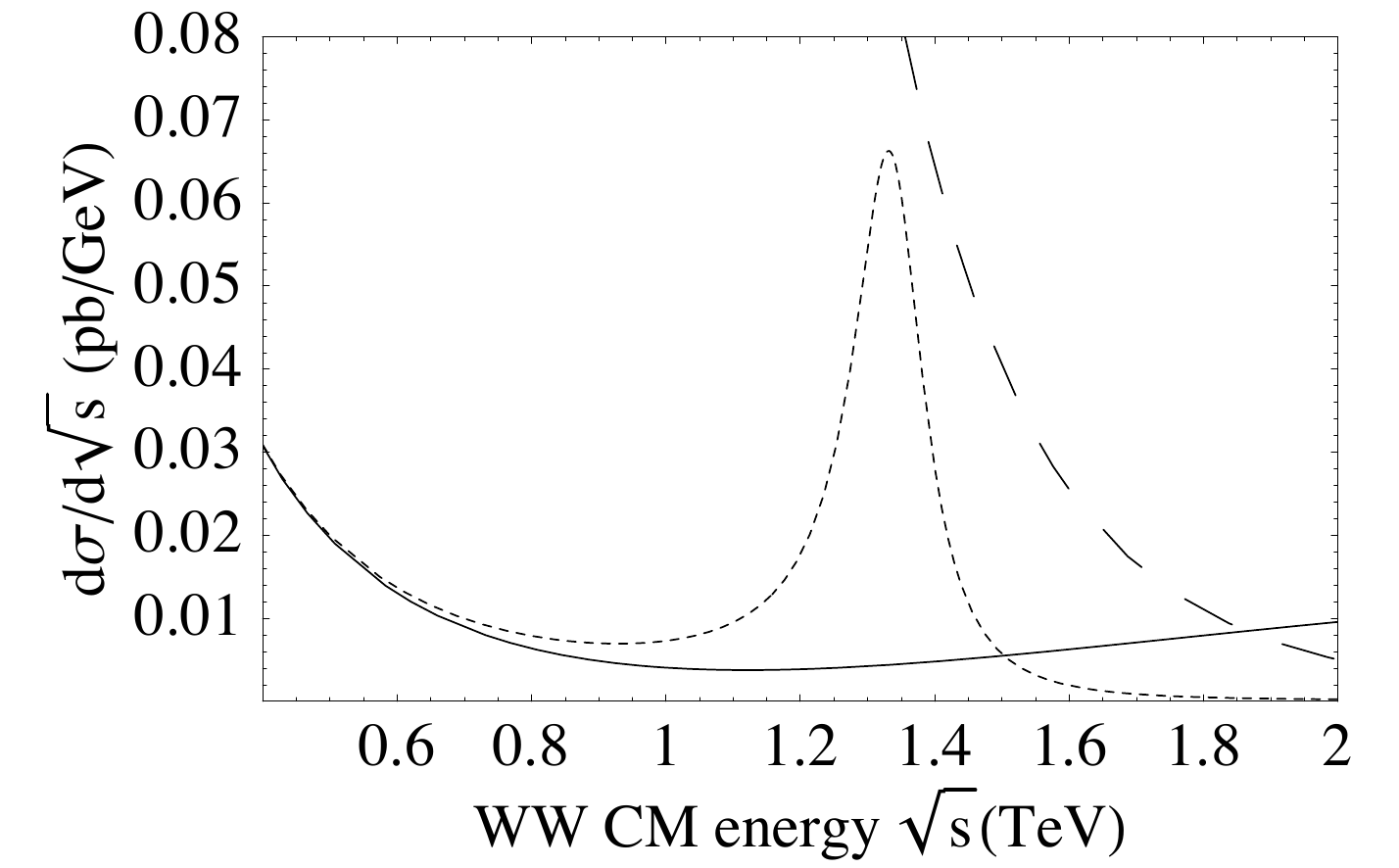}\includegraphics[width=3.5in]{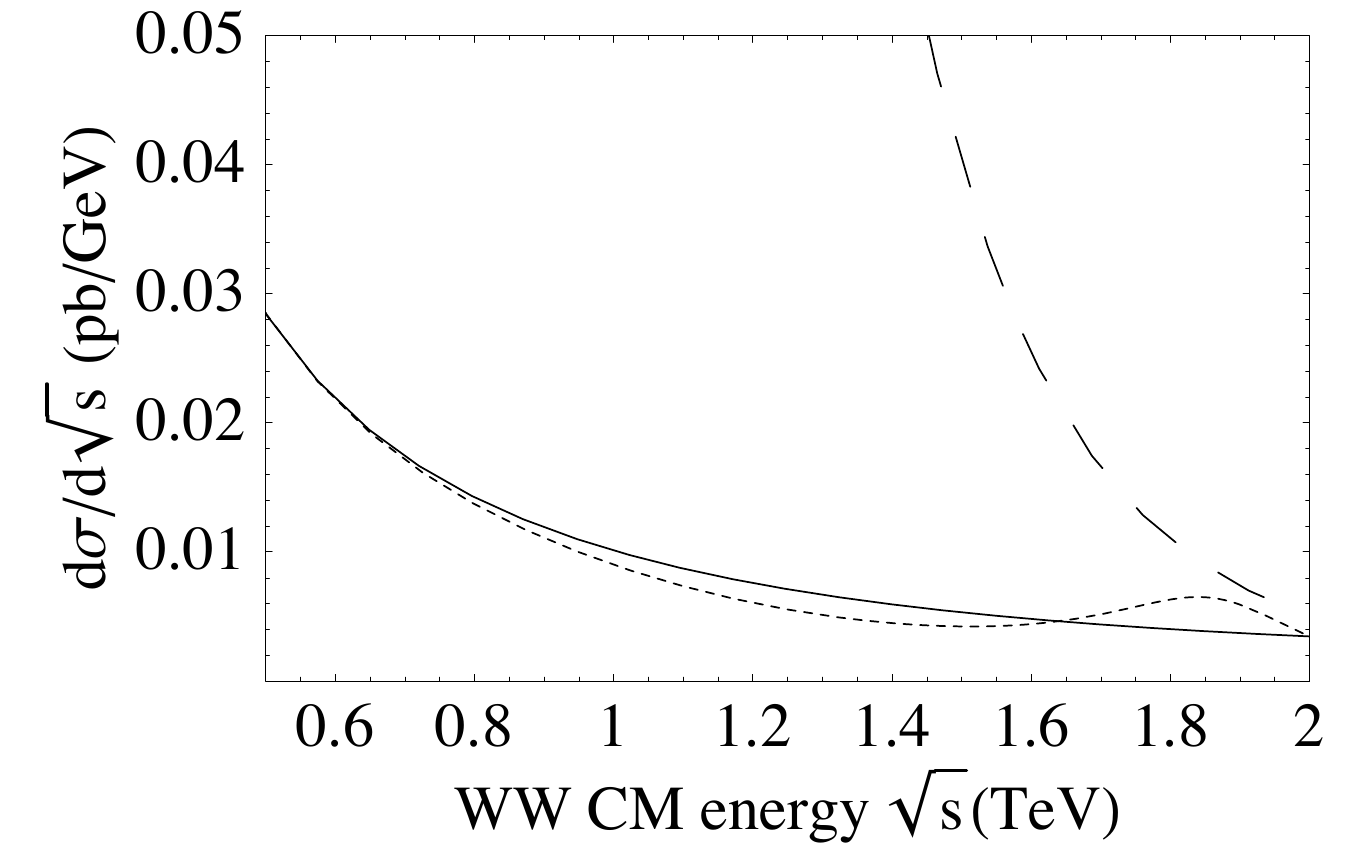} }
  \caption{\small  Parton-level cross sections for $WW$ scattering. In both figures, the continuous line is the result of the effective lagrangian. The long-dashed line is the  limit after which unitarity is lost. The dashed  line with a peak is the amplitude in presence of a vector resonance in the large-$N$ scenario.
  The two figures correspond to the  two representative points $P_1$ and $P_2$ discussed in the text. \label{fig3} }
 \end{figure}
\begin{figure}[h]
\includegraphics[width=3.5in]{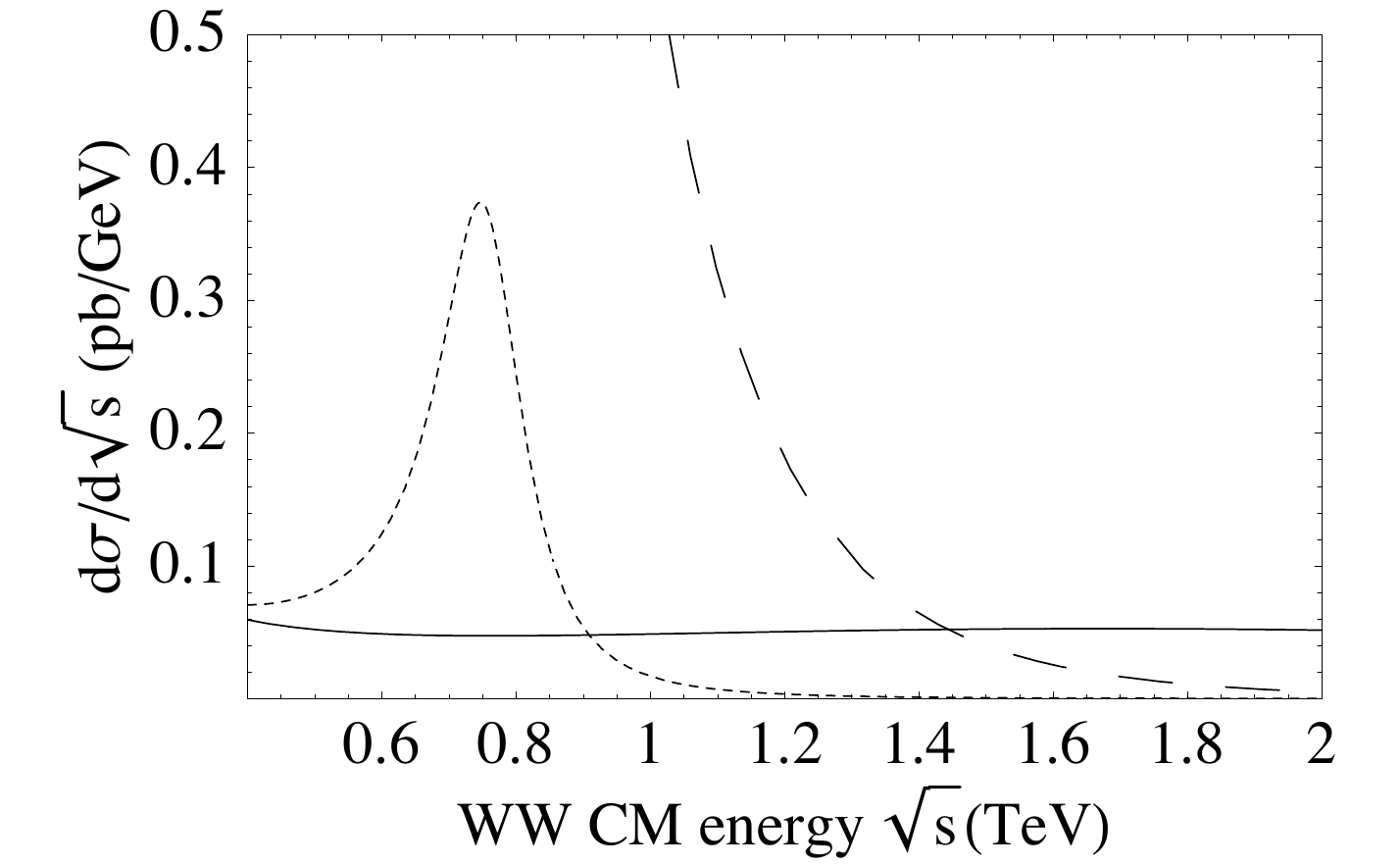}\includegraphics[width=3.5in]{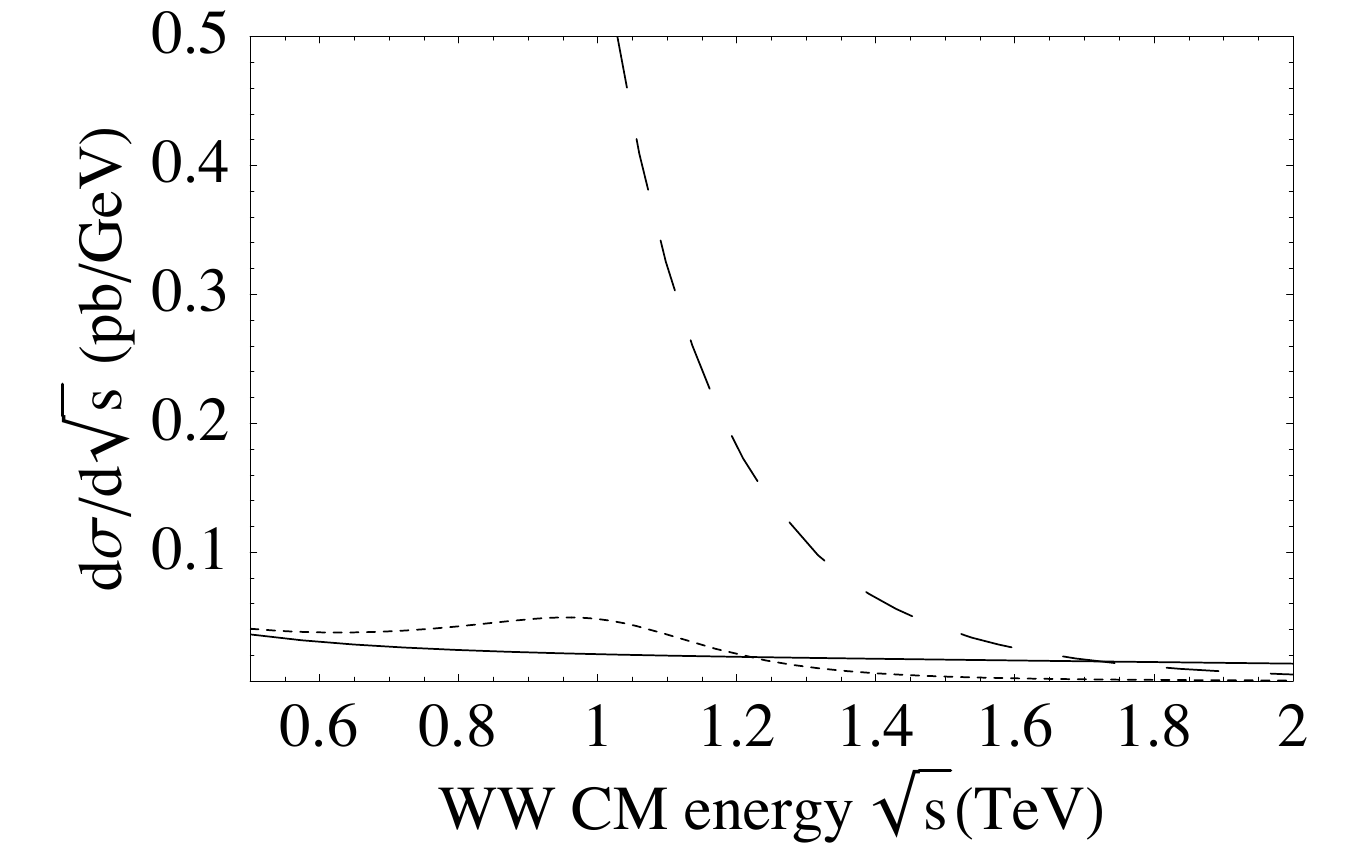}
\caption{\small  Parton-level cross sections for $WW$ scattering. The continuous line is the result of the effective lagrangian. The long-dashed line is the  limit after which unitarity is lost. The dashed line with a peak is the amplitude in presence of a scalar resonance in the heavy-Higgs scenario. The two figures correspond to the  two representative points $P_3$ and $P_4$ discussed in the text. \label{fig4}}
\end{figure}

We can now consider the physical process $pp\rightarrow W_LW_Ljj+X$ and plot its differential cross section in the $WW$ CM energy $\sqrt{s}$ for the values of the coefficients $a_4$ and $a_5$ we have identified. 
To simplify,  we will use the effective $W$ approximation~\cite{EWA}.

Once the amplitude $A(s,t,u)$ is given, the differential cross-section for the factorized $WW$ process is
\begin{equation}
\label{WWxsecn}
\frac{d\sigma_{WW} }{d\cos \theta }=\frac{|A(s,t,u)|^{2}}{32\pi \, s}.
\end{equation}
while the differential cross section for the considered physical transition $pp\rightarrow W_LW_Ljj+X$  reads:
\begin{equation}
\label{ppxsecn}
\frac{d\sigma }{ds}=\sum _{i,j}\int ^{1}_{s/s_{pp}}\int _{s/(x_{1}s_{pp})}^{1}\frac{dx_{1}\, dx_{2}}{x_{1}x_{2}s_{pp}}f_{i}(x_{1},s)\, f_{j}(x_{2},s)\frac{dL_{WW}}{d\tau }\int ^{1}_{-1}\frac{d\sigma_{WW}}{d\cos \theta }d\cos \theta 
\end{equation}
where \( \surd s_{pp} \) is the CM energy which we take to be 14
TeV, as appropriate for the LHC, and
\be
\frac{dL_{WW}}{d\tau }\approx \left( \frac{\alpha }{4\pi \sin ^{2}\theta _{W}}\right) ^{2}\frac{1}{\tau }\left[ (1+\tau )\ln (1/\tau )-2(1-\tau )\right] 
\ee
where \( \tau =s/(x_{1}x_{2}s_{pp}) \). For the structure functions $f_j$ we use those of ref.~\cite{lai}.

The high-energy regime will be very much suppressed by the partition functions so that the resonances found by (\ref{scalar0}) and (\ref{vector1}) turn out to be the only phenomenologically interesting ones. Because of this, we can safely make use of the approximation (\ref{IAM})  in the whole range from 400 GeV to 2 TeV and thus we  take $A(s,t,u)$ to be given by the IAM unitarization of (\ref{Wamp}). 

Figures~\ref{fig3} and \ref{fig4} give the cross section for the large-$N$ and heavy-Higgs scenario, respectively. The scalar resonance corresponding to $P_3$  is particularly high and narrow and a very good candidate for detection.
For a LHC luminosity of 100 fb$^{-1}$, it would yield  $10^6$ events after one year.  If it exists, it will appear as what we would have called the Higgs boson even though it is not a fundamental state and its mass is much heavier than that expected for the SM Higgs boson.

The actual signal at the LHC requires that the parton-level cross sections derived here  be included in a Montecarlo simulation (of the bremsstrahlung of the initial partons, QCD showers as well as of the final hadronization) and compared with the expected background and the physics of the detector. In the papers of ref.~\cite{dobado,bcf}  it has been argued that resonances in the range here considered can be effectively identified at the LHC.

\acknowledgments

 This work is
partially supported by  MIUR  and the RTN European  Program MRTN-CT-2004-503369.  


 \end{document}